# RNNSecureNet: Recurrent neural networks for Cyber security use-cases

Mohammed Harun Babu R, Vinayakumar R, Soman KP

Center for Computational Engineering and Networking (CEN), Amrita School of Engineering, Coimbatore, Amrita Vishwa Vidyapeetham, India

harishharunn@gmail.com

## Abstract

Recurrent neural network (RNN) is an effective neural network in solving very complex supervised and unsupervised tasks. There has been a significant improvement in RNN field such as natural language processing, speech processing, computer vision and other multiple domains. This paper deals with RNN application on different use cases like Incident Detection, Fraud Detection, and Android Malware Classification. The best performing neural network architecture is chosen by conducting different chain of experiments for different network parameters and structures. The network is run up to 1000 epochs with learning rate set in the range of 0.01 to 0.5.Obviously, RNN performed very well when compared to classical machine learning algorithms. This is mainly possible because RNNs implicitly extracts the underlying features and also identifies the characteristics of the data. This helps to achieve better accuracy.

## Keywords

Deep neural networks (DNNs), recurrent neural network (RNN), Cyber security, Android malware classification, incident detection, fraud detection.

## Introduction

In today's data world, malware is the common threat to everyone from big organizations to common people and we need to safeguard our systems, computer networks, and valuable data. Cyber-crimes have risen to the peak and many hacks, data stealing, and many more Cyber-attacks. Hackers gain access through any loopholes and steal all valuable data, passwords and other useful information. Mainly in android platform malicious attacks increased due to increase in large number of application. In other hand its very easy for persons to develop multiple malicious malwares and feed it into android market very easily using a third party software's. Attacks can be through any means like e-mails, exe files, software, etc. Criminals make use of security vulnerabilities and exploit their opponents. This forces the importance of an effective system to handle the fraudulent activities. But today's sophisticated attacking algorithms avoid being detected by the security mechanisms. Every day the attackers develop new exploitation techniques and escape from Anti-virus and Malware software's. Thus nowadays security solution companies are moving towards deep learning and machine learning techniques where the algorithm learns the underlying information from the large collection of security data itself and makes predictions on new data. This, in turn, motivates the hackers to develop new methods to escape from the detection mechanisms.

Malware attack remains one of the major security threats in Cyberspace. It is an unwanted program which makes the system behave differently than it is supposed to behave. The solutions provided by antivirus software against this malware can only be used as a primary weapon of resistance because they fail to detect the new and upcoming malware created using polymorphic, metamorphic, domain flux and IP flux. The machine learning algorithms were employed which solves complex security threats in more than three decades [1]. These methods have the capability to detect new malwares. Research is going at a high phase for security problems like Intrusion Detection Systems (IDS), Malware Detection, Information Leakage, etc. Fortunately, today's Deep Learning (DL) approaches have performed well in various long-standing AI challenges [2] such as natural language processing (NLP), computer vision, speech recognition. Recently, the application of deep learning techniques have been applied for various use cases of Cyber security [3], [4], [5], [6], [7], [8], [9], [10], [11], [12], [13], [14], [15], [16], [17], [35]. It has the ability to detect the Cyber-attacks by learning the complex underlying structure, hidden sequential relationships and hierarchical feature representations from a huge set of security data. In this paper, we are evaluating the efficiency of SVM and RNN machine learning algorithms for Cyber security problems. Cyber security provides a set of actions to safeguard computer networks, systems, and data.

This paper is arranged accordingly where related work are discussed in Section 2 the background knowledge of recurrent neural network (RNN) in Section 3. In Section 4 proposed methodologies including description, data set are discussed and at last results are furnished in Section 5. Section 6 is concluding with conclusion.

## 2. Related works

In this section related work for Cyber security use cases is discussed: Android Malware Classification (Task 1), Incident Detection (Task 2), and Fraud Detection (Task 3). The most commonly used approach for Malware detection in Android devices is the static and dynamic approach [18]. In the static approach, all the android permissions are collected by unpacking the application and whereas, in dynamic approach, the run-time execution attributes like system calls, network connections, power consumption, user interactions and efficient utilization of memory. Most of the commercial systems used today use both the static and dynamic approach. For low computational cost, resource utilization, time resource Static analysis is mainly preferred for Android devices. Meanwhile dynamic analysis has the advantage to detect metamorphic and polymorphic malware. [19] have evaluated the performance of traditional ML algorithms for malware detection on Android devices without using the API calls and permission as features. MalDozer proposed the use of API calls with deep learning approach to detect the Android malware and classify them accordingly [20]. [21] API calls contains schematic information which helps in understand the intention of the app indirectly without any user interface. Using embedding techniques at training phase API calls are extracted using DEX assembly [20] which helps in effective malware detection on neural networks.

The security issues in cloud computing are briefly discussed in [22]. [23] proposed ML-based anomaly detection that acts on the network, service and work-flow layers. A hybrid of both machine learning and rule based systems are combined for intrusion detection in the cloud infrastructure [24]. [25] shows how Incident Detection can perform well than intrusion detection. In [26] discusses a detailed study on 6 different traditional ML classifiers in finding the credit card frauds, financial frauds. Credit card frauds are detected using Convolution Neural Networks. Fraud Detection in crowd sourcing projects is discussed in [27].Statistical Fraud Detection method model is trained to discriminate the fraudulent and non-fraudulent using supervised and unsupervised methods in credit card frauds. [21]Especially in communication networks Fraud Detection are rectified using supervised learning by statistical learning of behavior of networks us using Bayesian network approach. Data mining approaches related to financial Fraud Detection are discussed in [28]. [29] mainly discusses the Fraud Detection in today's new Online e-commerce transaction using Recurrent Neural Network(RNN) which performed very well. Based on this a detailed survey is conducted in [30]. The risks and trust involved in e-commerce market are detailed studied in [31].

## 3. Experiments

### 3.1. Description of Data-sets

The first task, Task 1 is an Android classification task. The dataset is created from a set of APK packages files collected from the Opera Mobile Store from Jan to Sep 2014 is used. This dataset consists of API (Application Programming Interface) information for 61,730 APK files where 30,897 files for training and 30,833 files for testing [32]. The second task, Task 2 is incident detection. This dataset contains operational log file that was captured from Unified Threat Management (UTM) of UniteCloud. Task 3 is Fraud Detection. This dataset is anonymised data that was unified using the highly correlated rule based uniformly distributed synthetic data (HCRUD) approach by considering similar distribution of features. Detailed description of Task 2 and Task 3 is given in **Tables** 1 and 2 respectively.

| # Samples | #Features | #Classes | #Training | #Testing |
|---|---|---|---|---|
| 100,000 | 9 | 2 | 70,000 | 30,000 |

**Table 1** Description of Task 2 Data set

| # APK files | #Features | #Classes | #Training | #Testing |
|---|---|---|---|---|
| 100,000 | 12 | 3 | 70,000 | 30,000 |

**Table 2** Description of Task 3 Data set

### 3.2 Hyper-parameter selection

In order to identify suitable parameter for Recurrent Network, we used a moderately sized architecture with one hidden layer consisting of 64, 128, 256, 512, and 768 units. 3 trails of the experiment are run for each parameter related to units and each experiment is run till 400 epochs. 768 units have shown highest 10-fold cross-validation accuracy for all use cases of Cyber security. Hence we decided to use 768 units for the rest of the experiments. To find an optimal result, three trails of experiment with 700 epochs have run with learning rate varying in the range [0.01-0.5]. The highest 10-fold cross-validation accuracy was obtained by using the learning rate of 0.01. There was a sudden decrease in accuracy at learning rate 0.05 and finally attained highest accuracy at learning rates of 0.035, 0.045 and 0.05 in comparison to learning rate 0.01. This accuracy may have been enhanced by running the experiments till 1000 epochs. As more complex architectures we have experimented with, showed less performance within 500 epochs, so 0.01 as learning rate for the rest of the experiments by taking training time and computational cost into account.

### 3.3 Network topologies

The RNN 1 to 6 layer network topology are used in order to find an optimum network structure for our input data since we don't know the optimal number of layers and neurons. We run 3 trails of experiments for each RNN network toplogy. Each trail of the experiment was run till 700 epochs. It was observed that most of the deep learning architectures learn the normal category patterns of input data within 400 epochs itself. The number of epochs required to learn the malicious category data usually varies. This complex architecture networks required a large number of iterations in order to reach the best accuracy. At last, we obtained the best performed network topology for each use case. For Task 2 and Task 3, 3 layer RNN network performed well. For Task One, the 6 layer RNN network gave a good performance in comparison to the 4 layer RNN. Then we decided to use 6 layer RNN network for the rest of the experiments. 10-fold cross-validation accuracy of each RNN network topology for all use cases is shown in **Table 3**.

| RNN network topology | Task Name | Accuracy |
|---|---|---|
| RNN 1 layer | Task 1 | 0.512 |
| RNN 2 layer | Task 1 | 0.611 |
| RNN 3 layer | Task 1 | 0.624 |
| RNN 4 layer | Task 1 | 0.634 |
| RNN 5 layer | Task 1 | 0.691 |
| RNN 5 layer | Task 1 | 0.701 |
| RNN 1 layer | Task 2 | 0.612 |
| RNN 2 layer | Task 2 | 0.714 |

| RNN 3 layer | Task 2 | 0.827 |
| RNN 4 layer | Task 2 | 0.859 |
| RNN 5 layer | Task 2 | 0.896 |
| RNN 7 layer | Task 2 | 0.925 |
| RNN 1 layer | Task 3 | 0.611 |
| RNN 2 layer | Task 3 | 0.704 |
| RNN 3 layer | Task 3 | 0.754 |
| RNN 4 layer | Task 3 | 0.802 |
| RNN 5 layer | Task 3 | 0.812 |
| RNN 7 layer | Task 3 | 0.853 |

**Table 3** Summary of test results

### 3.4. Proposed Architecture

An intuitive overview of our proposed RNN architecture for all use cases is shown in **Fig 1**. This consists of the input layer with six hidden layers and an output layer. An input layer contains 4896 neurons for Task One, 9 neurons for Task Two and 12 neurons for Task Three. An output layer contains 2 neurons for Task One, 3 neurons for Task Two and 2 neurons for Task Three. The detailed structure and configuration of proposed RNN architecture are shown in **Table 3**. The neurons in input to hidden layer and hidden to output layer are fully connected. The proposed Recurrent Network is composed of recurrent layers, fully-connected layers, batch normalization layers and dropout layers.

**Recurrent layers:** It contains the recurrent units/neurons. The units have self-connection/loops. This helps to carry out the previous time step information for the future time step.

**Batch Normalization and Regularization:** To obviate overfitting and speed up the RNN model training, Dropout (0.001)[33] and Batch Normalization[34] was used in between fully-connected layers. A dropout removes neurons with their connections randomly. In our alternative architectures for Task 1, the recurrent networks could easily overfit the training data without regularization even when trained on large number samples.

**Classification:** For classification, the final fully connected layer follows sigmoid activation function for Task One and Task Two, softmax for Task Three. The fully connected layer absorb the non-linear kernel and sigmoid layer output zero

(benign) and output one (malicious), softmax provides the probability score for each class.

The prediction loss for Task 1 and Task 2 is estimated using binary cross entropy

$$loss(pd, ed) = -\frac{1}{N}\sum_{i=1}^{N}[ed_i \log pd_i + (1 - ed_i)\log(1 - pd_i)]$$

where $pd$ is a vector of predicted probability for all samples in testing data set, $ed$ is a vector of expected class label, values are either 0 or 1.

The prediction loss for Task 3 is estimated using categorical-cross entropy

$$loss(pd, ed) = -\sum_x pd(x)\log(ed(x))$$

where $pd$ is true probability distribution, $ed$ is predicted probability distribution. We have used $sgd$ as an optimizer to minimize the loss of binary-cross entropy and categorical-cross entropy.

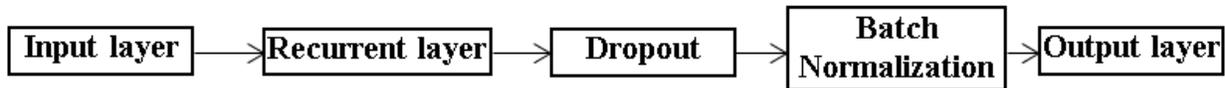

**Fig 1** Proposed architecture

## 4. Results

We have evaluated the proposed RNN model against classical machine learning classifier SVM, on 3 different Cyber security use cases. 1. Identifying Android malware based on API information, 2. Incident Detection over unified threat management (UTM) operation on Unite Cloud, 3. Fraud Detection in financial transactions. The detailed results of proposed RNN model on 3 different use cases are displayed in Table 4.

| Algorithm | Task Name | Accuracy | Precision | Recall | F-score |
|---|---|---|---|---|---|
| SVM | Android Malware Classification | 0.723 | 0.159 | 0.239 | 0.191 |
| SVM | Incident Detection | 0.993 | 0.998 | 0.992 | 0.995 |
| SVM | Fraud Detection | 0.916 | 0.922 | 0.916 | 0.917 |
| RNN 5 layer | Android Malware Classification | 0.741 | 0.098 | 0.215 | 0.134 |
| RNN 5 layer | Incident Detection | 0.997 | 0.999 | 0.998 | 0.998 |
| RNN 5 layer | Fraud Detection | 0.918 | 0.922 | 0.918 | 0.919 |

Table 4 Summary of test results

## 5. Conclusion

In this paper performance of RNN and other classical machine learning classifiers are evaluated for Cyber security use cases such as Android malware classification, incident detection, and fraud detection. In all the three use cases, RNN outperformed all the classical machine learning classifiers. Moreover, the same architecture for all three use cases is able to perform better than the other classical machine learning classifiers. The reported results of RNNs can be further improved by training with few more layers and neurons to the existing architectures.